# Precise Real-Time Navigation of LEO Satellites Using a Single-Frequency GPS Receiver and Ultra-Rapid Ephemerides

Xiucong Sun, Chao Han, Pei Chen [*]

*School of Astronautics, Beihang University, Beijing 100191, China*

**Abstract**

Precise (sub-meter level) real-time navigation using a space-capable single-frequency global positioning system (GPS) receiver and ultra-rapid (real-time) ephemerides from the international global navigation satellite systems service is proposed for low-Earth-orbiting (LEO) satellites. The C/A code and L1 carrier phase measurements are combined and single-differenced to eliminate first-order ionospheric effects and receiver clock offsets. A random-walk process is employed to model the phase ambiguities in order to absorb the time-varying and satellite-specific higher-order measurement errors as well as the GPS clock correction errors. A sequential Kalman filter which incorporates the known orbital dynamic model is developed to estimate orbital states and phase ambiguities without matrix inversion. Real flight data from the single-frequency GPS receiver onboard China's SJ-9A small satellite are processed to evaluate the orbit determination accuracy. Statistics from internal orbit assessments indicate that three-dimensional accuracies of better than 0.50 m and 0.55 mm/s are achieved for position and velocity, respectively.

*Keywords*: Precise real-time navigation; LEO; Single-frequency GPS receiver; Ultra-rapid ephemerides; Sequential Kalman filter

---

[*] Corresponding author. Tel.: +86 10 82316535.
E-mail address: chenpei@buaa.edu.cn (P. Chen).

# 1. Introduction

The global positioning system (GPS) receivers are widely used for orbit determination of low-Earth-orbiting (LEO) satellites nowadays. Centimeter level positioning accuracies have been achieved with geodetic dual-frequency (DF) GPS receivers [1-3]. Compared to DF receivers, single-frequency (SF) receivers are much cheaper and require lower energy consumption, and thus are more suitable for low-cost small satellites as well as space missions having relatively lower navigation accuracy requirements. The representative examples of space-capable SF receivers include the TOPSTAR 3000 receiver implemented on Korea multi-purpose satellite-2 (KOMPSAT-2), the MosaicGNSS receivers on TerraSAR-X/TanDEM-X, as well as the Phoenix receiver on the project for onboard autonomy-2 (PROBA-2) [4-7].

For SF GPS orbit determination, the ionospheric path delay cannot be eliminated by dual-frequency combination [8]. Bock et al. [9] compared three different methods of dealing with ionospheric effects, which are ionospheric modeling, linear combination, and parameter estimation. The linear combination method, i.e., the GRAPHIC (group and phase ionospheric correction) combination, showed the best performance for satellites at low altitudes. Three dimensional (3D) position accuracies of about 20 cm and 10 cm were achieved for the GRACE-A/B satellites. The GRAPHIC combination method was also employed for orbit determination of PROBA-2, and a 30 cm (3D) position accuracy was achieved [7].

The high accuracies mentioned above, i.e., centimeter for DF receivers and sub-meter for SF receivers, are usually obtained by a batch least-squares estimator in ground-based orbit determination using post-processed precise GPS ephemerides. In recent years, there has been an emerging requirement for accurate knowledge of orbit (sub-meter level) in real-time or near real-time space applications such as the onboard geocoding of high-resolution imagery, the open-loop operation of spaceborne altimeters, the atmospheric limb sounding and the maintenance of satellite constellations [10-14]. To support real-time precise positioning applications, the international global navigation satellite systems service (IGS) nowadays provides ultra-rapid products which contain 48 hours of ephemerides (the first half computed from observations with an accuracy of 3 cm and a latency of a few hours, and the second half predicted orbit with an accuracy of 5 cm). Compared to the satellite orbits, the accuracy of clock corrections is much lower (approximately 3 ns, corresponding to 0.9 m when multiplied by the speed of light). In order to satisfy real-time users with more precise products, the IGS established a real-time working group (RTWG) in 2001 and officially launched the real-time service (RTS) on April 1, 2013 [15]. These IGS real-time products could enable the real-time precise orbit determination (POD) for onboard processing which will further contribute to an increased autonomy of spacecraft operation in the future.

Efforts have to date been made to improve the accuracy of real-time orbit determination for LEO satellites. The Kalman filtering technique gradually gains popularity due to its recursive nature and the advantage of optimal combination of dynamical models and observation. In [16], the extended Kalman filter (EKF) is used to process DF code and phase data from SAC-C, and a 3D position error of 1.5 m is demonstrated with the use of broadcast ephemerides. In [10], a better accuracy of 0.31 m is achieved for SAC-C owing to the use of IGS ultra-rapid products. To further improve accuracy, the IGS ultra-rapid orbits combined with the real-time clock estimation (RETICLE) clocks developed by the German Space Operations Center were utilized for near real-time orbit determination of TerraSAR-X, and an orbit accuracy of better than 10 cm (3D) was obtained [13]. However, all these real-time POD experiments are based on DF GPS data. More recently, Want et al. [17] used the C/A code and L1 phase data from the BlackJack DF receiver onboard GRACE-A for real-time orbit determination test. A position accuracy of 0.55 m was achieved using the GRAPHIC combination method and broadcast ephemerides. In particular, Montenbruck et al. [7] processed the actual flight data from the Phoenix-XNS SF receiver onboard PROBA-2, and an in-flight accuracy of 1.1 m (3D) was achieved with broadcast ephemerides.

This study focuses on real-time POD of LEO satellites using a SF GPS receiver and IGS precise ephemerides. The single-differenced GRAPHIC combination strategy is employed to eliminate first-order ionospheric effects and receiver clock offsets simultaneously. The higher-order measurement errors which have been neglected in previous research, such as the instrumental delay [18] and the phase wind up [19], are listed in the observation equation and their effects are discussed. The IGS ultra-rapid products are used to provide real-time precise GPS satellite orbits and clock corrections, and their accuracies are evaluated by comparison with IGS final products. The IGS RTS products are not utilized since that the official data are not available for the test period in this study. The phase ambiguities are modeled as random-walk processes in order to incorporate the time-varying and satellite-specific higher-order measurement errors as well as the GPS clock correction errors. A sequential Kalman filter which incorporates the known orbital dynamic model is developed to estimate orbital states and phase ambiguities. The real-time orbit determination procedure is tested offline with actual flight data from the China's SJ-9A (Shi Jian-9A) low-cost small satellite [20]. Several internal orbit assessment methods have been used to evaluate the accuracy of the orbit determination results.

The structure of this paper is organized as follows. Section 2 introduces the single-differenced GRAPHIC observation model and analyzes the accuracy of IGS ultra-rapid products. Section 3 presents the stochastic orbit dynamic model, the measurement equation, the model configuration, and the sequential Kalman filter design. The experimental results are discussed in Section 4. Conclusions are drawn in Section 5.

## 2. Observation model

### 2.1. Single-differenced GRAPHIC observation

A spaceborne SF GPS receiver provides C/A code and L1 carrier phase measurements only

$$C_1 = \rho_r^s + c(\delta t_r - \delta t^s) - c\delta t_{rel} + I_{C_1} + d_{C_1} + gdv_{C_1} + M_{C_1} + \varepsilon_{C_1} \tag{1}$$

$$L_1 = \rho_r^s + c(\delta t_r - \delta t^s) - c\delta t_{rel} + A_r^s - I_{L_1} + d_{L_1} + \Delta\phi + pcv_{L_1} + M_{L_1} + \varepsilon_{L_1} \tag{2}$$

where $C_1$ and $L_1$ are the C/A code and L1 carrier phase in meters, respectively, $\rho_r^s$ the geocentric distance from receiver to satellite (w.r.t. antenna phase centers), $c$ vacuum speed of light, $\delta t_r$ and $\delta t^s$ the receiver and satellite clock offsets, $\delta t_{rel}$ the relativistic correction term, $A_r^s$ the constant phase ambiguity in meters, $I_{C_1}$ and $I_{L1}$ the ionospheric propagation delays, $d_{C_1}$ and $d_{L_1}$ the instrumental code and phase delays, $gdv_{C_1}$ the group delay variation [21,22], $\Delta\phi$ the phase wind-up correction, $pcv_{L_1}$ the phase center variation [23,24], $M_{C_1}$ and $M_{L_1}$ the multipath effects, and $\varepsilon_{C_1}$ and $\varepsilon_{L_1}$ are the measurement noises.

The GRAPHIC linear combination is defined as

$$\begin{aligned} G &\triangleq \frac{1}{2}(C_1 + L_1) \\ &= \rho_r^s + c(\delta t_r - \delta t^s) - c\delta t_{rel} + \delta I + \frac{1}{2}A_r^s + \frac{1}{2}(d_{C_1} + d_{L_1} + gdv_{C_1} + \Delta\phi + pcv_{L_1}) + M_G + \varepsilon_G \end{aligned} \tag{3}$$

where $\delta I = \dfrac{I_{C_1} - I_{L_1}}{2}$ is the ionospheric delay residual, $M_G$ represents $\dfrac{M_{C_1} + M_{L_1}}{2}$, and $\varepsilon_G = \dfrac{\varepsilon_{C_1} + \varepsilon_{L_1}}{2}$ represents the GRAPHIC observation noise. The standard deviation of $\varepsilon_G$ is

$$\sigma_G = \frac{1}{2}\sqrt{\sigma_{C_1}^2 + \sigma_{L_1}^2} \tag{4}$$

where $\sigma_{C_1}$ and $\sigma_{L_1}$ are the standard deviations of C/A code and L1 phase noises, respectively.

The GRAPHIC combination eliminates only the first-order ionospheric path delay. The higher-order ionospheric effects are on the order of centimeters [25]. The instrumental code and phase delays could be constant or time-varying [26]. The temporally constant but satellite-specific parts will be absorbed by the non-integer ambiguity, and the temporally variable but satellite independent parts will be absorbed by the receiver clock offset [27]. The same strategy can be applied to multipath effects as well as group delay and phase center variations, which are induced by the radiation patterns of the satellite and receiver antennas and vary as functions of the direction of the signals [21-24]. The remaining temporally variable and satellite-specific parts of the higher-order errors (the ionospheric residual, the instrumental delays, the multipath effects, the group delay variations, and the phase center variations) can be modeled as a random-walk process and will be absorbed by a time-varying phase ambiguity. Thus, the GRAPHIC observation model (3) is simplified as

$$G = \rho_r^s + c\left(\delta t_G - \delta t^s\right) - c\delta t_{rel} + \Delta\phi_G + A_G + \varepsilon_G \tag{5}$$

where $\delta t_G$ is the redefined receiver clock offset, $\Delta\phi_G$ denotes $\frac{\Delta\phi}{2}$, and $A_G$ is the redefined time-varying phase ambiguity and is modeled as a random-walk process

$$\dot{A}_G = \upsilon_G \tag{6}$$

where $\upsilon_G$ is a zero-mean Gaussian noise.

The single-differenced GRAPHIC observation is obtained by subtracting the GRAPHIC observations from two different GPS satellites

$$G^{ij} = \rho_r^{ij} - c\left(\delta t^{ij} + \delta t_{rel}^{ij}\right) + \Delta\phi_G^{ij} + A_G^{ij} + \varepsilon_G^{ij} \tag{7}$$

where the single-differencing (SD) operator is defined as $(*)^{ij} = (*)^j - (*)^i$, the superscripts $i, j$ denote the GPS satellites, and $A_G^{ij}$ is the time-varying SD GRAPHIC ambiguity. The receiver clock offset is eliminated by the SD operation.

In model (7), the SD geocentric distance term $\rho_r^{ij}$ implicitly contains the positions of receiver and GPS satellite antenna phase centers. The GPS satellite antenna phase center position and clock offsets are provided in the broadcast navigation message or the IGS orbit and clock products corrected with the absolute phase center offsets [23]. The relativistic and phase wind-up correction terms can be evaluated according to [19]. The noise $\varepsilon_G^{ij}$ is unpredictable but its statistical characteristic is usually known beforehand. The receiver antenna position and the SD GRAPHIC ambiguity remain as the unknowns to be resolved.

*2.2. Accuracy of IGS ultra-rapid products*

Model (7) does not consider the GPS ephemeris errors, which are an important concern in actual precise orbit determination of LEO satellites. The GPS ephemeris errors will introduce an additional error term in the SD GRAPHIC observation model

$$\Delta G^{ij} = \left(e_r^j \Delta r^j - e_r^i \Delta r^i\right) - c\left(\Delta \delta t^j - \Delta \delta t^i\right) \tag{8}$$

where $\Delta \boldsymbol{r}^i$ and $\Delta \boldsymbol{r}^j$ are position errors of the $i$th and $j$th GPS satellites, $\Delta \delta t^i$ and $\Delta \delta t^j$ are GPS clock correction errors, and $\boldsymbol{e}_r^i$ and $\boldsymbol{e}_r^j$ are the light-of-sight (LOS) vectors from the receiver to GPS satellites. The LOS vectors are close to the unit vectors of GPS satellites. Thus $\Delta G^{ij}$ can be approximated by

$$\Delta G^{ij} \approx \left(\Delta r^j - \Delta r^i\right) - c\left(\Delta \delta t^j - \Delta \delta t^i\right) \tag{9}$$

where $\Delta r^i$ and $\Delta r^j$ are the radial position errors of the $i$th and $j$th GPS satellites.

This section evaluates the accuracy of IGS ultra-rapid ephemerides by comparison with IGS final products, which have the highest quality (precision of 1-2 cm for both the GPS orbits and clocks) and can be used as reference [19]. The differences of GPS satellite radial positions and clock corrections between the ultra-rapid and final products are shown in Fig. 1. Three GPS satellites (PRN 5, 14, and 26) during the period from October 25, 2012 00:00:00 (GPS time) to October 25, 2012 23:59:59 (GPS time) are investigated. The Root-Mean-Squares (RMS) values of the GPS satellite radial position differences are quite small, which are 1.00 cm, 1.46 cm, and 0.86 cm, respectively. Thus the GPS orbit errors can be neglected in orbit determination. By contrast, the RMS values of clock correction differences (multiplied by *c*) are much larger, which are 0.46 m, 0.59 m, and 0.43 m, respectively. Similar phenomenon have been found for the other GPS satellites, although not presented here.

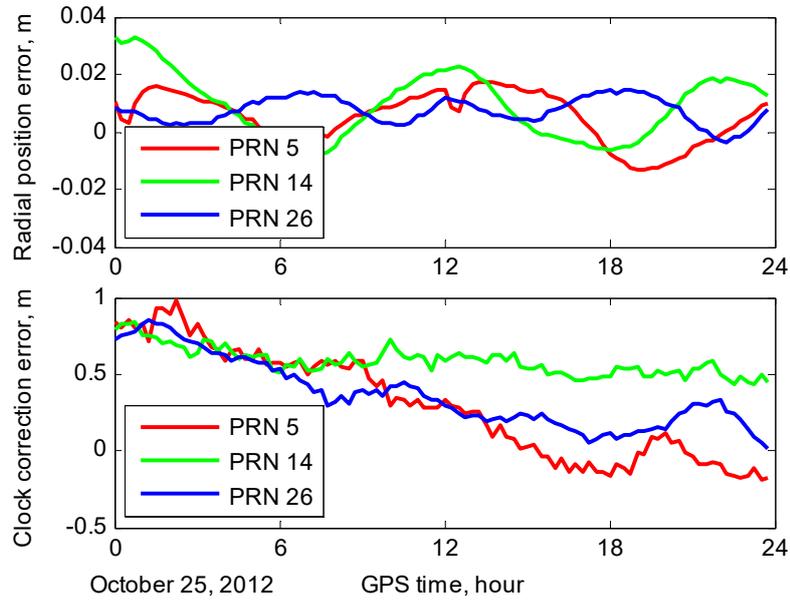

**Fig. 1** Accuracy of IGS ultra-rapid products using IGS final products as reference (15-min sampling interval).

The large clock correction errors would induce significant orbit determination errors if not addressed properly. Due to the GPS satellite frequent rising and setting, the typical visibility period of a GPS satellite as seen from a LEO user satellite is about 20-30 min. Sterle et al. [27] defined the concept of observation cluster as a set of carrier phase observations from a single satellite during a continuous time period that belongs to the same ambiguity. This concept can also be applied for the SD GRAPHIC observations. Therefore, the clock corrections are broken down into several pieces belonging to different observation clusters. The errors of ultra-rapid clock corrections for a 7-hour SJ-9A GPS tracking (divided into 3-5 clusters) is illustrated in Fig. 2. The time-varying error of each piece of clock corrections can be modeled as a slowly drifting bias contaminated with white noise and thus could be absorbed by the redefined random-walk ambiguity.

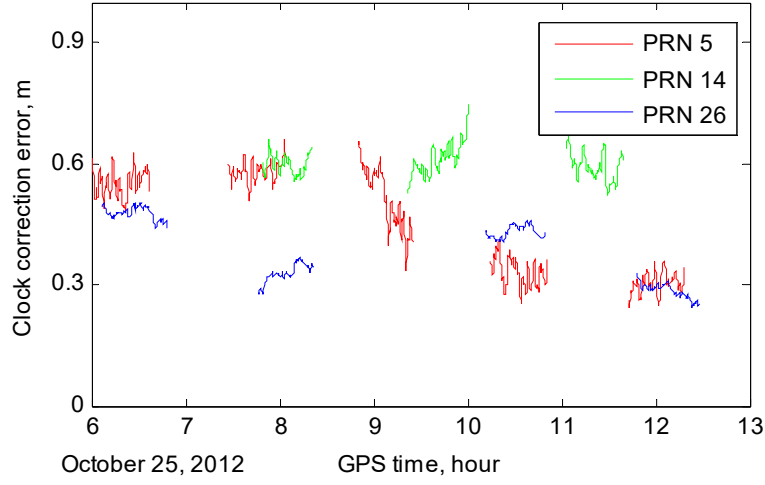

**Fig. 2** Errors of ultra-rapid clock corrections belonging to different observation clusters for SJ-9A (10-s sampling interval).

## 3. Real-time orbit determination

*3.1. Stochastic orbit modeling*

The SJ-9A satellite orbits at an altitude of about 650 km. The maximum area-to-mass ratio of SJ-9A is about 0.016 m$^2$/kg. The accelerations due to atmospheric drag and solar radiation pressure are on the orders of $1 \times 10^{-7}$ m/s$^2$ and $1 \times 10^{-8}$ m/s$^2$, respectively. Thus, the non-gravitational forces are not modeled in this study. The stochastic dynamic equation is given as follows

$$\dot{r} = v \tag{10}$$

$$\dot{v} = a_g + a_{s/m} + w_t \tag{11}$$

where $r$ and $v$ are the inertial position and velocity vectors of the satellite's center of mass (CoM), $a_g$ is the acceleration due to the Earth's gravity filed, $a_{s/m}$ is the acceleration due to solar and lunar gravitational attractions, and $w_t$ is a zero-mean white and Gaussian noise representing unmodeled accelerations.

The continuous time orbit dynamic model is discretized as follows

$$(r_k, v_k) = f(r_{k-1}, v_{k-1}, t_k, t_{k-1}) + w_{k-1} \tag{12}$$

where $f(\cdot)$ is a 6-dimensional vector function relating orbital states at epochs $t_k$ and $t_{k-1}$, and $w_{k-1}$ is the discrete process noise. The function $f(\cdot)$ has no explicit expression and is numerically obtained using an ordinary differential equation (ODE) solver.

The partial derivative of $(r_k, v_k)$ with respect to $(r_{k-1}, v_{k-1})$ is called the orbital state transition matrix and is obtained by integration of the following differential equation from $\tau = 0$ to $\tau = t_k - t_{k-1}$

$$\frac{d}{d\tau}\boldsymbol{\Phi}(t_{k-1}+\tau, t_{k-1}) = \begin{bmatrix} \boldsymbol{0}_{3\times 3} & \boldsymbol{I}_{3\times 3} \\ \left(\frac{\partial a_g}{\partial r} + \frac{\partial a_{s/m}}{\partial r}\right)_{3\times 3} & \boldsymbol{0}_{3\times 3} \end{bmatrix} \boldsymbol{\Phi}(t_{k-1}+\tau, t_{k-1}) \tag{13}$$

with the identity matrix as the initial value.

The covariance matrix of $w_{k-1}$ is denoted as $\boldsymbol{Q}_{r,v,k-1}$ and is obtained by numerically integrating the following differential equation from $t_{k-1}$ to $t_k$

$$\frac{d\boldsymbol{Q}_{r,v}}{d\tau} = \begin{bmatrix} \boldsymbol{0}_{3\times 3} & \boldsymbol{I}_{3\times 3} \\ \left(\frac{\partial a_g}{\partial r} + \frac{\partial a_{s/m}}{\partial r}\right)_{3\times 3} & \boldsymbol{0}_{3\times 3} \end{bmatrix} \boldsymbol{Q}_{r,v} + \boldsymbol{Q}_{r,v} \begin{bmatrix} \boldsymbol{0}_{3\times 3} & \boldsymbol{I}_{3\times 3} \\ \left(\frac{\partial a_g}{\partial r} + \frac{\partial a_{s/m}}{\partial r}\right)_{3\times 3} & \boldsymbol{0}_{3\times 3} \end{bmatrix} + \begin{bmatrix} \boldsymbol{0}_{3\times 3} & \boldsymbol{0}_{3\times 3} \\ \boldsymbol{0}_{3\times 3} & \boldsymbol{\sigma}_{w_t}\boldsymbol{\sigma}_{w_t}^T \end{bmatrix} \tag{14}$$

with zero initial values. $\sigma_{w_t}$ is the standard deviation of $w_t$.

*3.2. Measurement equation*

Consider a LEO user satellite tracking $n + 1$ common GPS satellites, the $n$ SD GRAPHIC observations compensated with GPS clock offsets, relativistic and phase wind-up corrections can be cast into the following vectorial model

$$\begin{aligned} \boldsymbol{z} &\triangleq \begin{bmatrix} G^{\kappa 1} + c\left(\delta t^{\kappa 1} + \delta t_{rel}^{\kappa 1}\right) - \Delta\phi_G^{\kappa 1} \\ G^{\kappa 2} + c\left(\delta t^{\kappa 2} + \delta t_{rel}^{\kappa 2}\right) - \Delta\phi_G^{\kappa 2} \\ \vdots \\ G^{\kappa n} + c\left(\delta t^{\kappa n} + \delta t_{rel}^{\kappa n}\right) - \Delta\phi_G^{\kappa n} \end{bmatrix} \\ &= \boldsymbol{h}(\boldsymbol{r}, \boldsymbol{b}) + \boldsymbol{\varepsilon} = \begin{bmatrix} \|\boldsymbol{r}+\boldsymbol{\delta}-\boldsymbol{r}^1\| - \|\boldsymbol{r}+\boldsymbol{\delta}-\boldsymbol{r}^\kappa\| \\ \|\boldsymbol{r}+\boldsymbol{\delta}-\boldsymbol{r}^2\| - \|\boldsymbol{r}+\boldsymbol{\delta}-\boldsymbol{r}^\kappa\| \\ \vdots \\ \|\boldsymbol{r}+\boldsymbol{\delta}-\boldsymbol{r}^n\| - \|\boldsymbol{r}+\boldsymbol{\delta}-\boldsymbol{r}^\kappa\| \end{bmatrix} + \boldsymbol{b} + \boldsymbol{\varepsilon} \end{aligned} \tag{15}$$

where $\boldsymbol{h}(\cdot)$ represents the measurement function, $\kappa$ is the reference GPS satellite, $\boldsymbol{\delta}$ is the receiver antenna phase center offset, $\boldsymbol{r}^i$ $(i = 1, 2, \ldots, n, \kappa)$ is the GPS satellite antenna phase center position, $\boldsymbol{b}$ is the time-varying random-walk ambiguity vector, and $\boldsymbol{\varepsilon}$ is the observation noise.

The random-walk ambiguity vector $\boldsymbol{b}$ contains the actual phase ambiguities, the satellite-specific higher-order measurement errors, as well as the GPS clock correction errors and is modeled as

$$\dot{\boldsymbol{b}} = \boldsymbol{w}_b \tag{16}$$

where $\boldsymbol{w}_b$ is a zero-mean white and Gaussian noise vector.

The partial derivative matrices of $\boldsymbol{z}$ with respect to $\boldsymbol{r}$ and $\boldsymbol{b}$ are

$$\boldsymbol{H}_r = \begin{bmatrix} \left(\boldsymbol{e}_r^\kappa - \boldsymbol{e}_r^1\right)^T \\ \left(\boldsymbol{e}_r^\kappa - \boldsymbol{e}_r^2\right)^T \\ \vdots \\ \left(\boldsymbol{e}_r^\kappa - \boldsymbol{e}_r^n\right)^T \end{bmatrix}, \quad \boldsymbol{H}_b = \boldsymbol{I}_{n \times n} \tag{17}$$

where $\boldsymbol{e}_r^i$ is the LOS vector from the receiver to the $i$th GPS satellite and $\boldsymbol{I}_{n \times n}$ is a $n \times n$ unity matrix.

The observation noise $\boldsymbol{\varepsilon}$ is assumed to be white and Gaussian. The covariance matrix of $\boldsymbol{\varepsilon}$ is

$$\boldsymbol{R} = \begin{bmatrix} 2 & 1 & \cdots & 1 \\ 1 & 2 & \cdots & 1 \\ \vdots & \vdots & \ddots & \vdots \\ 1 & 1 & \cdots & 2 \end{bmatrix} \sigma_G^2 \tag{18}$$

where $\sigma_G$ is the standard deviation of the GRAPHIC observation noise.

*3.3. Model configuration*

Table 1 summarizes the overall dynamic and measurement models for SJ-9A real-time orbit determination. The SD GRAPHIC observations and the IGS ultra-rapid ephemerides (predicted) provide the measurement information. The transmitter and receiver antenna phase center offsets and the relativistic and phase wind-up corrections are applied. The Earth Gravitational Model 2008 (EGM2008) [28,29] truncated at degree and order 40 is used to compute the accelerations due to the Earth's static gravity field. The third body attractions are computed using the analytical formulas of the solar and lunar positions [30]. The higher degree and order coefficients of the Earth's gravity model, the tidal effects, as well as the non-gravitational forces are not modeled and are included into the process noise. The standard deviation of the process noise should reflect the accuracy of the dynamic models and is set to $1 \times 10^{-6}$ m/s$^2$. The fourth-order Runge-Kutta with Richardson extrapolation is used as the ODE solver. The computation of these models refers to the ICRF/ITRF2008 coordinate system [29].

Table 1  Summary of dynamic and measurement models used in real-time orbit determination of SJ-9A

| Item | SJ-9A real-time POD |
| --- | --- |
| GPS measurement model | Ionosphere-free and receiver-clock-offset-removal L1 code-carrier combination (SD GRAPHIC) |
| | IGS ultra-rapid GPS ephemerides and clocks |
| | Phase center offsets of transmitter and receiver antennas |
| | Relativistic and phase wind-up corrections |
| | Ambiguities with random-walk noise of 1 cm |
| | Elevation cut-off 5˚, sampling rate 10 s |
| Gravitational forces | EGM2008 model (40 × 40) |
| | No tidal models |
| | Solar and lunar gravitational attractions (analytical formulas) |
| Non-gravitational forces | No atmospheric drag and solar radiation pressure model |
| | White noise for unmodeled accelerations, $1 \times 10^{-6}$ m/s$^2$ |
| Numerical integration | Fourth-order Runge-Kutta with Richardson extrapolation |
| Reference frames | ICRF/ITRF2008 |

*3.4. Sequential Kalman filter design*

The Kalman filter consists of two alternating stages of time update and measurement update. At the measurement update stage, the computation of filter gain requires the inversion of an $n \times n$ matrix, where $n$ is the number of measurements. Instead of processing $n$ measurements simultaneously, the sequential Kalman filter (SKF) implements measurement update one measurement at a time [31]. Therefore, the SKF avoids matrix inversion and has advantages for embedded navigation systems.

The detailed algorithm of the SKF for linear system estimation is provided in [31]. This section presents the overall filter design for the SF GPS obit determination. The state vector comprises the inertial position and velocity of the satellite's CoM and the SD GRAPHIC ambiguity vector

$$\boldsymbol{x}^T = \begin{bmatrix} \boldsymbol{r}^T & \boldsymbol{v}^T & \boldsymbol{b}^T \end{bmatrix} \quad (19)$$

The estimate of $\boldsymbol{x}$ is denoted by $\hat{\boldsymbol{x}}$ and the accompanying covariance is denoted by $\boldsymbol{P}$.

Given initial values of $\hat{\boldsymbol{x}}$ and $\boldsymbol{P}$, the filter processes single-differenced GRAPHIC measurements at consecutive epochs and recursively updates the state estimates and the covariance. The initial values of $\hat{\boldsymbol{r}}$ and $\hat{\boldsymbol{v}}$ are obtained by a 3rd-order polynomial fitting of short-arc (3 min) code-derived positions. The initial values of $\hat{\boldsymbol{b}}$ are obtained from code-minus-carrier observations. Conservative values of (10 m)², (0.5 m/s)², and (10 m)² are set for the covariances of initial position, velocity, and ambiguity estimates, respectively.

At the time-update stage, the filter states are propagated from previous epoch $t_{k-1}$ to current epoch $t_k$

$$\left(\hat{r}_k^-, \hat{v}_k^-\right) = f\left(\hat{r}_{k-1}^+, \hat{v}_{k-1}^+, t_k, t_{k-1}\right) \tag{20}$$

$$\hat{b}_k^- = \hat{b}_{k-1}^+ \tag{21}$$

$$\left(\hat{\bar{x}}_k^-\right)^T = \left[\left(\hat{r}_k^-\right)^T \quad \left(\hat{v}_k^-\right)^T \quad \left(\hat{b}_k^-\right)^T\right] \tag{22}$$

Here, we assume that the visible GPS satellites remains unchanged. The covariance is propagated as follows

$$P_k^- = F_{k,k-1} P_{k-1}^+ F_{k,k-1}^T + Q_{k-1} \tag{23}$$

with

$$F_{k,k-1} = \begin{bmatrix} \Phi(t_k, t_{k-1}) & 0_{6 \times n} \\ 0_{n \times 6} & I_{n \times n} \end{bmatrix} \tag{24}$$

$$Q_{k-1} = \begin{bmatrix} Q_{r,v,k-1} & 0_{6 \times n} \\ 0_{n \times 6} & Q_{b,k-1} \end{bmatrix} \tag{25}$$

The orbital state transition matrix $\Phi(t_k, t_{k-1})$ and the orbital dynamic process noise matrix $Q_{r,v,k-1}$ are obtained by numerical integration of Eqs. (13) and (14), respectively. The ambiguity process noise matrix $Q_{b,k-1}$ is given by

$$Q_{b,k-1} = \text{cov}(w_b)(t_k - t_{k-1}) \tag{26}$$

where $\text{cov}(w_b)$ is the covariance matrix of $w_b$.

At the measurement-update stage, the measurements are first decorrelated by linear transformation and are then processed one by one to update the state estimates. The measurement noise matrix $R_k$ is symmetric positive definite and can be decomposed as

$$R_k = S_k^T D_k S_k \tag{27}$$

where $S_k$ is an orthogonal matrix, and $D_k$ is a diagonal matrix

$$D_k = \begin{bmatrix} 1 & \cdots & 0 & 0 \\ \vdots & \ddots & \vdots & \vdots \\ 0 & \cdots & 1 & 0 \\ 0 & \cdots & 0 & n_k + 1 \end{bmatrix} \tag{28}$$

where $n_k$ is the number of tracked GPS satellites at epoch $t_k$. $\boldsymbol{S}_k$ and $\boldsymbol{D}_k$ can be precomputed offline to reduce the computational effort. The $n_k$ measurements are decorrelated by defining a new set of measurements

$$\tilde{\boldsymbol{z}}_k = \boldsymbol{S}_k \boldsymbol{z}_k \tag{29}$$

Initialize the a posteriori estimate and covariance as

$$\hat{\boldsymbol{x}}_{0k}^+ = \hat{\boldsymbol{x}}_k^-, \quad \boldsymbol{P}_{0k}^+ = \boldsymbol{P}_k^- \tag{30}$$

For $i = 1, \ldots, n$, perform the following measurement-update equations

$$\boldsymbol{K}_{ik} = \frac{\boldsymbol{P}_{i-1,k}^+ \tilde{\boldsymbol{H}}_{ik}^T}{\tilde{\boldsymbol{H}}_{ik} \boldsymbol{P}_{i-1,k}^+ \tilde{\boldsymbol{H}}_{ik}^T + D_{ik}} \tag{31}$$

$$\hat{\boldsymbol{x}}_{ik}^+ = \hat{\boldsymbol{x}}_{i-1,k}^+ + \boldsymbol{K}_{ik}\left(\tilde{z}_{ik} - \boldsymbol{S}_{ik}\boldsymbol{h}\left(\hat{\boldsymbol{r}}_{i-1,k}^+, \hat{\boldsymbol{b}}_{i-1,k}^+\right)\right) \tag{32}$$

$$\boldsymbol{P}_{ik}^+ = \left(\boldsymbol{I} - \boldsymbol{K}_{ik}\tilde{\boldsymbol{H}}_{ik}\right)\boldsymbol{P}_{i-1,k}^+\left(\boldsymbol{I} - \boldsymbol{K}_{ik}\tilde{\boldsymbol{H}}_{ik}\right)^T + \boldsymbol{K}_{ik}D_{ik}\boldsymbol{K}_{ik}^T \tag{33}$$

where $\hat{\boldsymbol{r}}_{i-1,k}^+$ and $\hat{\boldsymbol{b}}_{i-1,k}^+$ are the position and ambiguity components of $\hat{\boldsymbol{x}}_{i-1,k}^+$, $\tilde{z}_{ik}$ is the $i$th element of $\tilde{\boldsymbol{z}}_k$, $D_{ik}$ is the $i$th diagonal element of $\boldsymbol{D}_k$, $\boldsymbol{S}_{ik}$ is the $i$th row of $\boldsymbol{S}_k$, and $\tilde{\boldsymbol{H}}_{ik}$ is the $i$th row of the following Jacobian matrix

$$\tilde{\boldsymbol{H}}_k = \boldsymbol{S}_k \boldsymbol{H}_k = \boldsymbol{S}_k \begin{bmatrix} \boldsymbol{H}_{r,k} & \boldsymbol{0}_{n_k \times 3} & \boldsymbol{H}_{b,k} \end{bmatrix} \tag{34}$$

The a posterior estimate and covariance is assigned as

$$\hat{\boldsymbol{x}}_k^+ = \hat{\boldsymbol{x}}_{n_k,k}^+, \quad \boldsymbol{P}_k^+ = \boldsymbol{P}_{n_k,k}^+ \tag{35}$$

Before the measurement-update stage, an additional check of the change in observed GPS satellites is required. A reordering operation will be implemented if new GPS satellites are available, old satellites disappear, and the reference satellite changes [32].

After the check of satellite change, a simple but effective quality control procedure dealing with code outliers and/or carrier phase cycle slips is applied to guarantee robustness against erroneous measurements. The detecting is based on a channel-wise hypothesis test [32]. An outlier or cycle slip is considered to be present on the $i$th channel at epoch $k$ if

$$\left|z_{ik} - h_i\left(\hat{\boldsymbol{r}}_k^-, \hat{\boldsymbol{b}}_k^-\right)\right| > \sqrt{3\left(\boldsymbol{H}_{ik}\boldsymbol{P}_k^-\boldsymbol{H}_{ik}^T + R_{ik}\right)} \tag{36}$$

where $\hat{r}_k^-$ and $\hat{b}_k^-$ are the position and ambiguity components of $\hat{x}_k^-$, $z_{ik}$ is the $i$th element of $z_k$, $h_i(\cdot)$ is the $i$th component of $h(\cdot)$, $H_{ik}$ is the $i$th row of $H_k$, and $R_{ik}$ is the $i$th diagonal element of $R_k$. The detecting method can be viewed as a simple variant of that presented in Teunissen [33] with a moving widow of $N=1$. The channel-wise checking avoids matrix inversion and is suitable for real-time application. After detection of erroneous measurements, the corresponding channels will be isolated. The corresponding ambiguities will be reinitialized from code-minus-carrier observations and then be estimated at the following measurement-update stage.

In order to identify whether an outlier or a cycle slip occurs, the measurement residual at epoch $k+1$ is further involved. A cycle slip is identified if

$$\left| z_{i,k+1} - h_i\left(\hat{r}_{k+1}^-, \hat{b}_{k+1}^-\right) \right| < \sqrt{3\left(H_{i,k+1} P_{k+1}^- H_{i,k+1}^T + R_{i,k+1}\right)} \tag{37}$$

is satisfied. In this case, no further adapting is needed at epoch $k+1$. An outlier at epoch $k$ is identified if (37) is not satisfied but the following condition

$$\left| z_{i,k+1} - h_i\left(\hat{r}_{k+1}^-, \hat{b}_k^-\right) \right| < \sqrt{3\left(H_{i,k+1} P_{k+1}^- H_{i,k+1}^T + R_{i,k+1}\right)} \tag{38}$$

is satisfied. The condition (38) says that the estimated ambiguity at epoch $k-1$ is also valid at epoch $k+1$, which implies that an instantaneous outlier has occurred at epoch $k$. Then the ambiguity estimate $\hat{b}_{k+1}^-$ should be recovered to $\hat{b}_k^-$. If neither (37) or (38) is satisfied, a new event of outlier/cycle slip is considered to be present at epoch $k+1$, and the data quality control procedure will be repreated.

The flowchart of the SKF for real-time SF GPS orbit determination is depicted in Fig. 3. The filter starts with the initialization of the state vector and the covariance and then enters cycles of time-update and measurement-update stages. At the time-update stage, the position and velocity are propagated using the orbital dynamic model, and the ambiguities are directly passed to the current epoch. At the measurement-update stage, the SD GRAPHIC observations are decorrelated and sent to the filter one by one to update the state vector and covariance. The reordering operation and the quality control procedure are implemented between the two changes in order to handle situations where the sequence of observed GPS satellites changes and outliers/cycle slips occur.

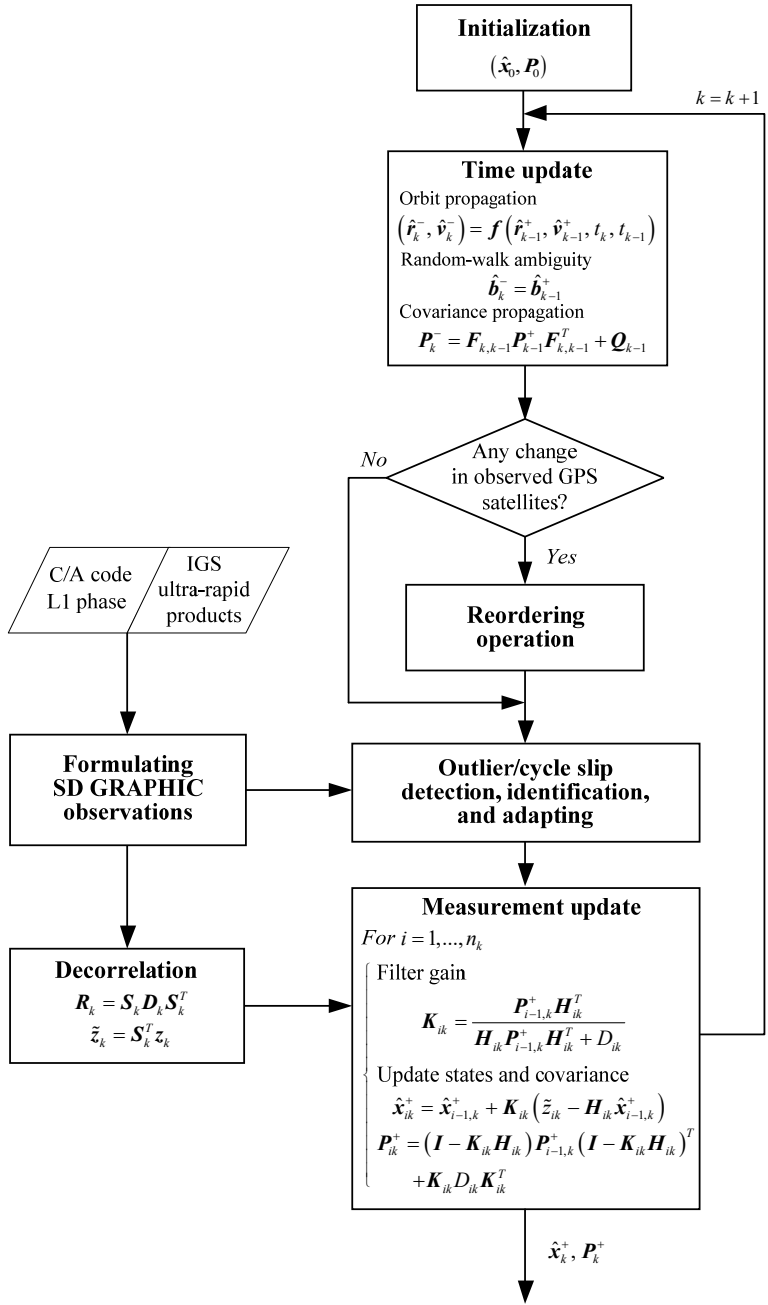

**Fig. 3** Flowchart of the sequential Kalman filter for real-time SF GPS orbit determination

## 4. Experimental results

A real-time orbit determination experiment based on actual flight data from the single-frequency receiver onboard the SJ-9A satellite has been carried out in an offline desktop computer environment. The SJ-9A is a fully maneuverable small satellite which performs a formation flying mission together with a micro-satellite SJ-9B [20]. Both the two satellites are manufactured by the China Academy of Space Technology (CAST). The SJ-9A is based on the mature CAST-2000 platform and the satellite mass is about 790 kg. The SJ-9B is based on the new CAST-100 platform and the mass is about 260 kg. Each satellite is equipped with two 12-channel single-frequency GPS receivers connected to a common antenna. A zero-baseline test between the two GPS receivers is conducted to roughly determine the standard deviations of the C/A code and L1 phase noises, which are 0.6 m and 1 mm, respectively. The receiver antenna phase center and the attitude information of SJ-9B are not provided. Thus, only the GPS data from SJ-9A are used for the test.

A 7-hour dataset referring to October 25, 2012 starting from 06:00:00 (GPS time) to 13:00:00 (GPS time) from the main GPS receiver onboard SJ-9A is used for the orbit determination test. During this time period, the SJ-9A satellite was flying in a sun-synchronous orbit with an altitude of about 650 km and an inclination of 98°, and the propulsion system was not working. The IGS ultra-rapid ephemeris file, igu17114_00.sp3, which was released at 03:00:00 (UTC time) and contains 24-hour predicted GPS ephemerides and clock corrections starting from 00:00:00 (GPS time) to 23:59:59 (GPS time) is used. The ephemeris and clock corrections errors are shown in Fig. 1. The measurements are processed at a sampling rate of 10 s. A cut-off elevation angle of 5° is utilized to reject GPS satellites in poor geometry. The standard deviation of the ambiguity's random-walk noise is set to 1 cm.

The sequential Kalman filter has been implemented by the orbit determination software developed by the Spacecraft Simulation Technology (SST) laboratory at Beihang University. The post-fit measurement residuals are first investigated to check the consistency of the dynamic model with the GPS data. As depicted in Fig. 4, the mean and standard deviation of the SD GRAPHIC residuals are -0.023 m and 0.472 m, respectively. The mean is close to zero, revealing the unbiasedness of the filter. The standard deviation of the residuals is slightly larger than the standard deviation of the SD GRAPHIC observation noise (0.431 m), revealing that the residuals include not only the observation noise but also the higher-order measurement errors.

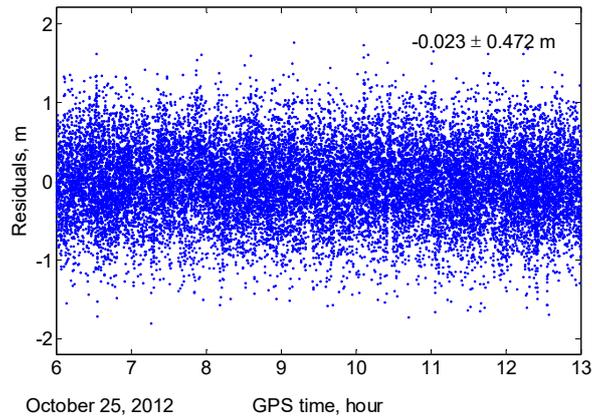

**Fig. 4** Post-fit SD GRAPHIC measurement residuals

The orbit overlap comparison provides a useful way to evaluate the internal consistency of orbit solutions. A backward sequential Kalman filter starting from the last epoch is implemented for the orbit overlap analysis. The forward and backward filters utilize GPS measurements before and after a given epoch and provide overlap information for state estimation at the current epoch. The orbit overlap differences are shown in Fig. 5. The filter convergence time is about 25 min. The RMS values of the radial, along-track, cross-track, and 3D position differences (after convergence) are 0.399 m, 0.476 m, 0.297 m, and 0.688 m, respectively. The RMS values of the radial, along-track, cross-track, and 3D velocity differences are 0.390 mm/s, 0.404 mm/s, 0.334 mm/s, and 0.653 mm/s, respectively. The RMS values of the overlap differences are roughly a factor of 1.414 higher than those of the absolute estimation errors (concluded from simulation analysis). Thus the accuracy of the SJ-9A real-time orbit determination experimental results can be deduced: 0.486 m for position and 0.462 mm/s for velocity.

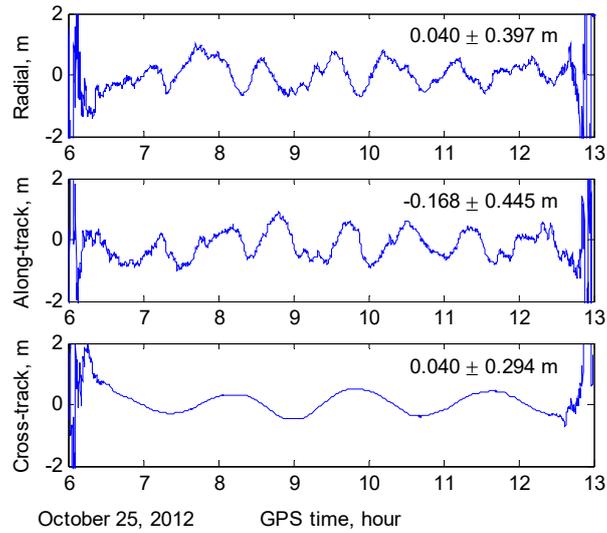

(a) Position difference

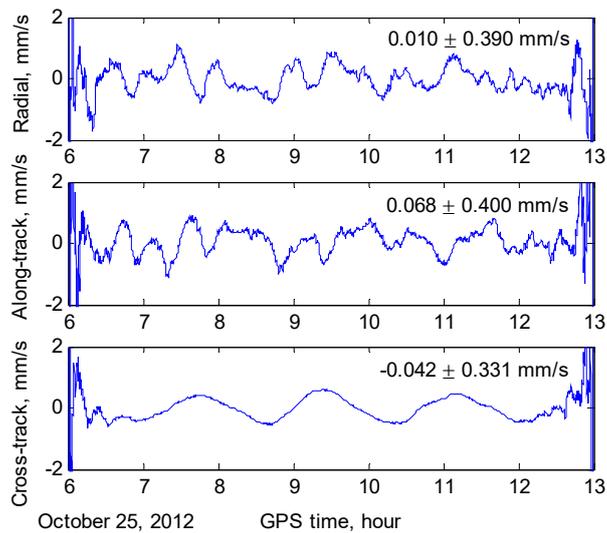

(b) Velocity difference

**Fig. 5** Orbit overlap comparison between the forward and backward filters

The real-time orbit determination experimental results are further compared with those obtained using a batch processing methodology. The batch orbit determination filter also processes SD GRAPHIC observations but utilizes the IGS final GPS ephemeris and clock products as reference. In addition, a more accurate orbital dynamic model is employed. As shown in Table 2, the EGM2008 gravity model truncated at degree and order 100 are used to compute the accelerations due to the Earth's static gravity field. The solid Earth and ocean tides are modeled [29]. The third body attractions are computed using the JPL Planetary and Lunar Ephemerides DE405 [34]. Piecewise constant empirical accelerations with a constraint of $1 \times 10^{-7}$ m/s$^2$ are used to represent the unmodeled accelerations and are estimated along with the orbital states. Orbit overlap analysis has demonstrated a 3D accuracy of 0.20 m for position and 0.11 mm/s for velocity.

Table 2  Dynamic models used in batch orbit determination of SJ-9A

| Item | SJ-9A batch POD |
| --- | --- |
| Gravitational forces | EGM2008 model (100 × 100) |
|  | Solid Earth and ocean tides (IERS2010) |
|  | Solar and lunar gravitational attractions (JPL DE405) |
| Non-gravitational forces | No atmospheric drag and solar radiation pressure model |
|  | Piecewise constant empirical accelerations at 15 min intervals, constraint: $1 \times 10^{-7}$ m/s$^2$ |

The differences between the real-time experimental results and the batch POD solutions are shown in Fig. 6. The RMS values of the radial, along-track, cross-track, and 3D position differences (after convergence) are 0.274 m, 0.284 m, 0.221 m, and 0.452 m, respectively. The RMS values of the radial, along-track, cross-track, and 3D velocity differences are 0.423 mm/s, 0.182 mm/s, 0.230 mm/s, and 0.515 mm/s, respectively. By removing the contributions of the batch POD errors, the 3D RMS position and velocity errors of the real-time experimental results are deduced to be 0.405 m and 0.503 mm/s.

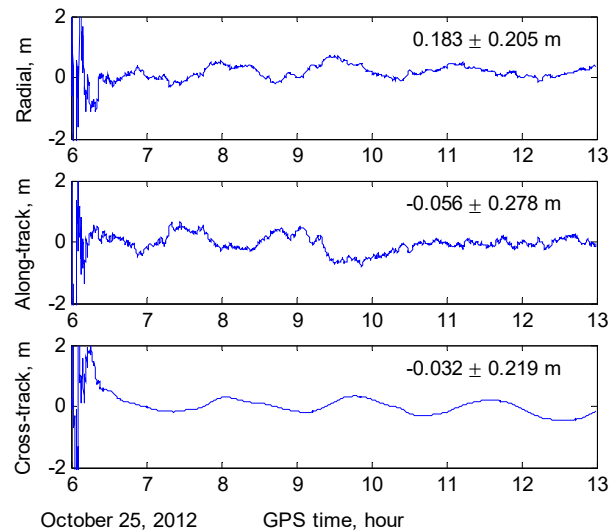

(a) Position difference

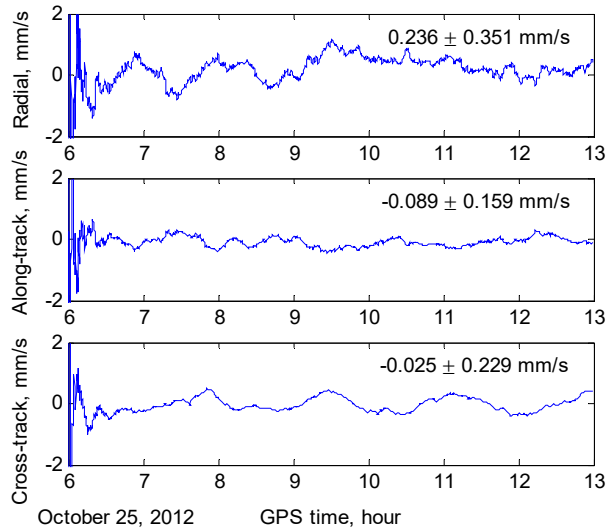

(b) Velocity difference

**Fig. 6** Comparison between real-time experimental results and batch orbit determination solutions

## 5. Conclusions

This study investigates real-time precise navigation of LEO satellites using a single-frequency GPS receiver and the predicted portion of IGS ultra-rapid products. A sequential Kalman filter which processes single-differenced GRAPHIC observations is developed and tested with actual flight data from the single-frequency GPS receiver onboard the SJ-9A small satellite. Internal consistency analysis indicates a position accuracy of better than 0.50 m (3D RMS) and a velocity accuracy of better than 0.55 mm/s (3D RMS).

The use of single-frequency GPS receivers is well compatible with real-time onboard space operations requiring sub-meter orbit accuracy and can reduce the hardware cost for LEO satellite tracking. Future study will be focused on further improvement of the real-time orbit determination accuracy with IGS RTS products.

## Acknowledgements

This research was supported by the National Natural Science Foundation of China through cooperative agreement No. 11002008 and has been funded in part by Ministry of Science and Technology of China through cooperative agreement No. 2014CB845303.

**Author Biographies**

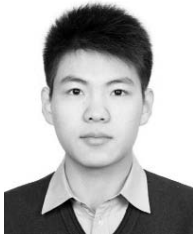

**Xiucong Sun** received his B.S. degree in aerospace engineering from Beihang University, Beijing, in 2010. He is currently a Ph.D. student at the school of astronautics, Beihang University. The focus of his current research mainly lies in spacecraft orbit and attitude determination, GNSS navigation.

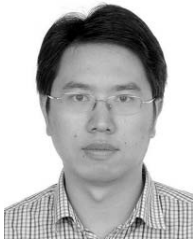

**Pei Chen** received his Ph.D. degree in aerospace engineering from Beihang University, Beijing, in 2008. He is currently an associate professor at the school of astronautics, Beihang University. His current research activities comprise spacecraft navigation, GNSS application, and astrodynamics and simulation.

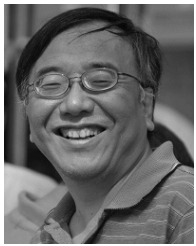

**Chao Han** received his M.S. and Ph.D. degrees in Applied Mechanics from Beihang University in 1985 and 1989, respectively. He is currently a professor at the school of astronautics, Beihang University. He is also an editorial board member of Chinese Journal of Aeronautics. The focus of his research activities lies in the area of spacecraft orbit and attitude dynamics, spacecraft guidance, navigation, and control.